# Multiplane Quantitative Phase Imaging Using a Wavelength-Multiplexed Diffractive Optical Processor

Che-Yung Shen[1,2,3], Jingxi Li[1,2,3], Tianyi Gan[1,3], Yuhang Li[1,2,3], Langxing Bai[4], Mona Jarrahi[1,3] and Aydogan Ozcan[1,2,3*]

[1]Electrical and Computer Engineering Department, University of California, Los Angeles, CA, 90095, USA

[2]Bioengineering Department, University of California, Los Angeles, CA, 90095, USA

[3]California NanoSystems Institute (CNSI), University of California, Los Angeles, CA, 90095, USA

[4]Department of Computer Science, University of California, Los Angeles, CA, 90095, USA

[*]Correspondence to: ozcan@ucla.edu

**Abstract**

Quantitative phase imaging (QPI) is a label-free technique that provides optical path length information for transparent specimens, finding utility in biology, materials science, and engineering. Here, we present quantitative phase imaging of a 3D stack of phase-only objects using a wavelength-multiplexed diffractive optical processor. Utilizing multiple spatially engineered diffractive layers trained through deep learning, this diffractive processor can transform the phase distributions of multiple 2D objects at various axial positions into intensity patterns, each encoded at a unique wavelength channel. These wavelength-multiplexed patterns are projected onto a single field-of-view (FOV) at the output plane of the diffractive processor, enabling the capture of quantitative phase distributions of input objects located at different axial planes using an intensity-only image sensor. Based on numerical simulations, we show that our diffractive processor could simultaneously achieve all-optical quantitative phase imaging across several distinct axial planes at the input by scanning the illumination wavelength. A proof-of-concept experiment with a 3D-fabricated diffractive processor further validated our approach, showcasing successful imaging of two distinct phase objects at different axial positions by scanning the illumination wavelength in the terahertz spectrum. Diffractive network-based multiplane QPI designs can open up new avenues for compact on-chip phase imaging and sensing devices.

# 1 INTRODUCTION

Quantitative phase imaging (QPI) stands as a powerful label-free technique capable of revealing variations in optical path length caused by weakly scattering samples[1–3]. QPI enables the generation of high-contrast images of transparent specimens, which are difficult to observe using conventional bright-field microscopy. In recent years, various QPI methodologies have been established, including e.g., off-axis imaging methods[4,5], phase-shifting methods[6,7], and common-path QPI techniques[8,9]. These methods have been instrumental in conducting precise measurements of various cellular dynamics and metabolic activities covering applications in, e.g., cell biology[10,11], pathology[12–14] and biophysics[15], such as the monitoring of real-time cell growth and behavior,[16,17] cancer detection[18,19], pathogen sensing[20,21], and the investigation of subcellular structures and processes[22]. In addition, QPI also finds applications in materials science and nanotechnology, which include characterizing thin films, nanoparticles and fibrous materials, revealing their unique optical and physical attributes[23–25].

Predominantly, QPI systems are employed to extract quantitative phase information within a two-dimensional (2D) plane by utilizing a monochromatic light source and sensor array. Given that standard optoelectronic sensors are limited to detecting only the intensity of light, advanced approaches utilizing customized illumination schemes and interferometric techniques[26–28], combined with digital post-processing and reconstruction algorithms, are employed to convert the intensity signals into quantitative phase images. Building on the foundations of 2D QPI approaches, tomographic QPI and optical diffraction tomography methods have also expanded QPI's capabilities to encompass volumetric imaging[29–32]. These techniques typically capture holographic images from multiple illumination angles, which allows for the digital reconstruction of the refractive index distribution across the entire 3D volume of the sample.

The digital post-processing techniques in QPI and phase tomography systems have witnessed a paradigm shift, primarily attributed to the recent advancements in the field of artificial intelligence. Specifically, the efficiency of feed-forward neural networks utilizing the parallel processing power of Graphics Processing Units (GPUs) has markedly increased the speed and throughput of image reconstruction in QPI systems[33–36]. These deep learning-based approaches facilitated solutions to various complex tasks of QPI, such as segmentation and classification[37–39], as well as inverse problems including phase retrieval[35,36,40–45], aberration correction[46,47], depth-of-field extension[48,49] and cross-modality image transformations[14,50]. Additionally, deep learning-based techniques have also been used to enhance 3D QPI systems by improving the accuracy and resolution of 3D refractive index reconstructions, utilizing methods such as physical approximant-guided learning[51], recurrent neural networks[52], neural radiance fields[53], alongside the reduction of coherent noise through generative adversarial networks[54]. However, the complexity of digital neural networks employed in these reconstruction techniques requires substantial computational resources, leading to lower imaging frame rates and increased hardware costs and computing power. These challenges become further intensified in 3D QPI systems due to the necessity of processing a larger set of interferometric images for 3D reconstructions.

Here, we introduce an all-optical, wavelength-multiplexed QPI approach that utilizes diffractive processing of coherent light to obtain the quantitative phase distributions of multiple phase objects distributed at varying axial depths. As illustrated in **Figure 1,** our approach employs a diffractive optical processor that is composed of spatially-engineered dielectric diffractive layers, optimized collectively via deep learning[55–65]. Following the deep learning-based design phase, these diffractive elements are physically fabricated to perform task-specific modulation of the incoming optical waves, converting the phase profile of each of the phase-only objects located at different axial planes into a distinct intensity distribution at a specific wavelength within its output field-of-view (FOV). These wavelength-multiplexed intensity distributions can then be recorded, either simultaneously with a multi-color image sensor equipped with a color filter array or sequentially using a monochrome detector by scanning the illumination wavelength to directly reveal the object phase information through intensity recording at the corresponding wavelength.

Based on this framework, we conducted analyses through numerical simulations and proof-of-concept experiments. Initially, we examined how the overlap of input objects at different axial positions affects the quality of the diffractive output images and all-optical quantitative phase information retrieval. Our results demonstrated that this diffractive QPI framework could achieve near-perfect quantitative phase imaging for phase objects without spatial overlap along the optical axis. Furthermore, even when the input objects are entirely overlapping along the axial direction, our diffractive processor could effectively reconstruct the quantitative phase information of each input plane with high fidelity and minimal crosstalk among the imaging channels. Beyond numerical analyses, we also experimentally validated our approach by designing and fabricating a diffractive multiplane QPI processor operating at the terahertz spectrum. Our experimental results closely aligned with the numerical simulations, confirming the practical feasibility of diffractive processors in retrieving the quantitative phase information of specimens across different input planes.

The presented diffractive multiplane QPI design incorporates wavelength multiplexing and passive optical elements, enabling the rapid capture of quantitative phase images of specimens across multiple axial planes. This system's notable compactness, with an axial dimension of < 60 mean wavelengths ($\lambda_\mathrm{m}$) of the operational spectral band, coupled with its all-optical phase recovery capability, sets it apart as a competitive analog alternative to traditional digital QPI methods. Additionally, the scalable nature of our design allows its adaptation to different parts of the electromagnetic spectrum by scaling the feature size of each diffractive layer proportional to the illumination wavelength of interest. Our presented framework paves the way for the development of new phase imaging solutions that can be integrated with focal plane arrays operating at various wavelengths to enable efficient, on-chip imaging and sensing devices, which can be especially valuable for applications in biomedical imaging/sensing, materials science, and environmental analysis, among others.

## 2 RESULTS

**Design of a wavelength-multiplexed diffractive processor for multi-plane QPI**

**Figure 1** presents a diagram of our diffractive multiplane QPI design that is based on wavelength multiplexing. In this setup, multiple transparent samples, which are axially separated, are illuminated by a broadband spatially coherent light. This broadband illumination can be regarded as a combination of plane waves at distinct wavelengths $\{\lambda_1, \lambda_2, ..., \lambda_M\}$, organized in order from the longest to the shortest wavelength. Here, $M$ represents the number of spectral channels as well as the number of phase objects/input planes, as each wavelength channel is uniquely assigned to a specific input plane. The illumination fields, denoted as $s_w$ ($w \in \{1,2,...,M\}$), propagate through multiple phase-only transmissive objects, each exhibiting a unique phase profile $\{\Psi_w\}$ at the corresponding input plane $P_w$. As the illumination light encounters the sample at each plane, it undergoes a phase modulation of $e^{j\Psi_w}$, resulting in multispectral optical fields $\{i_w\}$ at the input aperture of the diffractive processor. The wavelength-multiplexed QPI diffractive processor consists of several modulation layers constructed by dielectric materials, where each layer is embedded with spatially designed diffractive features that have a lateral size of $\sim\lambda_M/2$ with a trainable/optimizable thickness, covering a phase modulation range of 0 to $2\pi$ for all the illumination wavelengths. These diffractive layers, along with the input and output planes, are interconnected through optical diffraction in free space (air).

The complex fields $i_w$, resulting from the stacked input planes along the axial (z) direction, are modulated by the diffractive optical processor to yield output fields $\{o_w\}$, i.e., $o_w = \mathfrak{D}\{i_w\}$. The intensity variations of these output fields are then captured by a monochrome image sensor, which sequentially records the QPI signals across the illumination wavelengths. The resulting optical intensity measurements at each illumination wavelength, noted as $D_w$, can be expressed as:

$$D_w = |o_w|^2 \qquad (1).$$

Considering that the optical intensity $D_w$ recorded by the sensor is influenced by both the power of the illumination and the output diffraction efficiency, we used a straightforward normalization approach[61,66] to counteract potential fluctuations caused by power variations and achieve consistent QPI performance. This involves dividing the output measurements ($D_w$) into two zones: an output signal area $\mathcal{S}$, and a reference signal area $\mathcal{R}$. Here, $\mathcal{R}$ is designated as a one-pixel wide border surrounding the edges of $D_w$. This border is further segmented into $M$ subsections, each labeled as $\mathcal{R}_w$ ($w \in \{1,2,...,M\}$). A given $\mathcal{R}_w$ acts as a reference signal ($Ref_w$) for the wavelength channel $\lambda_w$, i.e.,

$$Ref_w = \frac{1}{N^{(\mathcal{R}_w)}} \sum_{(x,y)\in\mathcal{R}_w} D_w(x,y) \qquad (2),$$

where $N^{(\mathcal{R}_w)}$ denotes the total number of image sensor pixels located within $\mathcal{R}_w$. Finally, the output quantitative phase image $\{\Phi_w\}$ of the wavelength multiplexed diffractive processor can be obtained through a simple normalization step:

$$\Phi_w = \frac{D_w}{Ref_w} \qquad (3).$$

Once the training of our diffractive multiplane QPI processor successfully converges, all the output quantitative phase images $\boldsymbol{\Phi}_w$ obtained at different wavelengths $\lambda_w$ are expected to approximate the phase profiles of the input objects $\boldsymbol{\Psi}_w(\lambda_w)$, which can be written as:

$$\boldsymbol{\Phi}_w \approx \boldsymbol{\Phi}_w^{(GT)} = \boldsymbol{\Psi}_w(\lambda_w) \quad (4).$$

Here the ground truth phase images $\boldsymbol{\Phi}_w^{(GT)}$ are defined, without loss of generality, as the object phase distributions $\boldsymbol{\Psi}_w(\lambda_w)$ at the corresponding wavelength $\lambda_w$. Based on the above formulation, our diffractive multiplane QPI processor is optimized to act as an all-optical transformer that simultaneously performs two tasks:

(1) a space-to-spectrum transformation that encodes spatial information of input objects at different axial positions into different spectral channels, and

(2) a phase-to-intensity transformation that converts phase information of input objects into intensity distributions within the output FOV. This approach facilitates the reconstruction of multiplane quantitative phase information using solely a single output FOV.

To optimize/train our diffractive multiplane QPI processor, we compiled a training dataset of 110,000 images containing 55,000 handwritten images and 55,000 custom-designed grating/fringe-like patterns[67]. During training, to form each multiplane object, $M$ images were randomly chosen from these 110,000 training images without replacement and encoded into the phase channels ($\boldsymbol{\Psi}_w$) of the $M$ object planes. For this phase encoding, we adopted an assumption that all the object planes are composed of the same material and an identical range of material thickness variations. This assumption ensures that these different object planes, regardless of their individual axial positions, induce a similar magnitude of phase modulations on the incoming complex fields, thereby better mirroring the real-world scenarios encountered in multiplane imaging systems. Based on this assumption, we chose to have the thickness profile $h(x, y)$ of each phase-only object plane to be confined within the same dynamic range of $[0, H_{\text{tr}}\lambda_M]$, where $H_{\text{tr}}$ stands for the thickness range parameter used during the training, defined based on the shortest wavelength $\lambda_M$. Following this formulism, for the $w^{\text{th}}$ object plane, the maximum phase modulation $\varphi_{\text{tr},w}$ of the incoming field at wavelength $\lambda_w$ can be written as:

$$\varphi_{\text{tr},w} = \frac{2\pi}{\lambda_w}(n_o(\lambda_w) - 1)H_{\text{tr}}\lambda_M \quad (5),$$

where $n_o(\lambda_w)$ denotes the refractive index of the object material at $\lambda_w$. Accordingly, we also define a phase contrast parameter $\alpha_{\text{tr},w} = \frac{\varphi_{\text{tr},w}}{\pi}$ to represent the maximum phase contrast of objects at wavelength $\lambda_w$, i.e.,

$$\alpha_{\text{tr},w} = \frac{2}{\lambda_w}(n_o(\lambda_w) - 1)H_{\text{tr}}\lambda_M \quad (6).$$

As a result, the phase modulation values in each object plane are confined to a range of $[0, \alpha_{\text{tr},w}\pi]$. Without loss of generality, in our numerical analyses, we chose $H_{\text{tr}} = 0.6$ and a constant material refractive index ($n_o$) of 1.5 for all $\lambda_w$. As a result, the phase contrast parameter values $\alpha_{\text{tr},w}$ vary according to the operational wavelength, peaking at the shortest wavelength $\lambda_M$, where $\alpha_{\text{tr},M} = H_{\text{tr}}$ = 0.6. Error-backpropagation and stochastic gradient descent were employed to optimize the thickness of the diffractive layers by minimizing a custom loss function $\mathcal{L}$ defined based on the mean-squared error (MSE) between the diffractive output quantitative phase images and their ground truth across all the wavelength channels, i.e., $\mathcal{L} = \frac{1}{N_w}\sum_{w=1}^{N_w} MSE(\boldsymbol{\Phi}_w^{(GT)}, \boldsymbol{\Phi}_w)$. More information about the training process is provided in the Method section.

To numerically demonstrate the feasibility of our diffractive system, we devised several diffractive multiplane QPI processors, focusing on the impact of input object lateral overlap—where the fields of view of the input objects located at different axial planes overlap in the x and y directions. The occurrence of lateral overlap, resulting in nonuniform illumination, can deteriorate the quality of QPI reconstructions. To explore the dynamics between adjacent input phase objects during image reconstruction and assess our design's capability in handling laterally overlapping objects at different axial planes, we adapted our training models to various assumptions about the lateral separation between different axial planes. These five input phase objects were uniformly distributed on the circumference of a circle with a radius of r from the center, as shown in **Figure S1, Supporting Information**. A maximum lateral separation distance R was set as ~94.5$\lambda_\text{m}$, ensuring that the input FOVs are not distributed out of the boundary of a diffractive layer. Building on this, we developed and trained six distinct diffractive designs by adjusting the lateral separation distance (r) of the input planes across various values spanning {0, 0.2R, 0.4R, 0.6R, 0.8R, R}, as illustrated in **Figure 2a**. These different configurations of input object arrangements, which cover the condition of a complete spatial overlap (r = 0) of objects to a complete lateral separation (r = R), enabled us to investigate the impact of r on the system's QPI performance. Apart from the varying input lateral separations, these diffractive multiplane QPI designs share identical input specifications, featuring the same number of input planes $M = 5$ with $N_x^{(\Psi)} \times N_y^{(\Psi)} = 14 \times 14$. All the diffractive designs are composed of 10 diffractive layers, where each diffractive layer has 600 × 600 trainable diffractive features. The entire diffractive volume spans an axial length of ~56.2 $\lambda_\text{m}$ and a lateral size of ~262.5$\lambda_\text{m}$, forming a compact system that can be monolithically integrated with a CMOS image sensor. At the output plane of these diffractive designs, a monochrome image sensor with a pixel size of ~5.2$\lambda_\text{m} \times 5.2\lambda_\text{m}$ is assumed. A unit magnification is selected between the object/input plane and the monochrome output/sensor plane, resulting in the same size of the output signal region $\mathcal{S}$ as the input FOV for each axial plane. After their deep learning-based optimization, the thickness profiles of the diffractive layers for each of the six designs are depicted in **Figure S2, Supporting Information**.

**Performance analysis of wavelength-multiplexed diffractive processors for multi-plane QPI**

After the training stage, we first conducted blind testing of the resulting diffractive processor designs through numerical simulations. To evaluate the multiplane QPI performance of these designs, we constructed a test set comprising 5,000 phase-only objects, which were never used in the training process. These objects were synthesized by randomly selecting images from the MNIST dataset and encoding them into the phase channels of the $w^\text{th}$ input object with a dynamic phase range of $[0, \alpha_{\text{test},w}\pi]$. Mirroring the approach used during the training, the phase ranges in the testing were derived from a thickness range of [0, $H_{\text{test}}\lambda_M$], consistent across the $M$ input planes, where $H_{\text{test}}$ stands for the testing thickness range parameter. The corresponding diffractive QPI output examples of the blind testing results are visualized in **Figure 2b**. Here, the Pearson Correlation Coefficient (PCC) was utilized to quantify the performance of these diffractive processor designs.

From the observation of the output examples with $H_{\text{test}} = H_{\text{tr}} = 0.6$ shown in **Figure 2b**, it is evident that a large lateral separation distance (r = R) among different axial planes ensures a decent reconstruction of inputs, yielding high-fidelity output images. Conversely, a smaller lateral separation distance, such as r = 0.4R, results in diminished image contrast and the introduction of some imaging artifacts. We noted a consistent degradation in the image quality as the input lateral separation distance r was reduced from R to 0. This decrease in the QPI performance can be attributed to two main factors:

(1) unknown sample-induced nonuniform illumination from the neighboring input planes in front of the target axial plane, and

(2) phase disturbance when propagating through the axially stacked input planes after the target planes.

When the testing thickness range is larger than the training thickness range, i.e., $H_{\text{test}} = 1 > H_{\text{tr}}$, the output QPI results of r = 0 were found to be degraded. These output images can hardly be recognized because of stronger phase perturbations caused by the larger phase contrast at each object plane. On the contrary, the output QPI measurements of r = R still present a good image fidelity at the output of the wavelength multiplexed diffractive QPI processor for $H_{\text{test}} = 1 > H_{\text{tr}}$. These results highlight the diffractive design's ability to process and image phase-only objects with a larger thickness and higher phase contrast beyond what was encountered during the training phase, i.e., $H_{\text{test}} > H_{\text{tr}}$.

We also evaluated the resulting PCC values in **Figure 3**, which reflects the examples shown in **Figure 2b**. As revealed in **Figure 3a**, the design with complete lateral separation of inputs (r = R) achieved high output PCC scores across all the imaging channels when $H_{\text{tr}} = H_{\text{tr}} = 0.6$, reaching an average PCC value of 0.993 ± 0.001, corroborating the observations from visual inspections. When the input phase objects were completely laterally overlapping (r = 0), the output PCC values dropped to 0.884 ± 0.016, whereas the reconstructed digit images could still be discernable. When $H_{\text{test}}$ increased to 1, as shown in **Figure 3b**, the performance of the design with complete lateral separation of inputs (r = R) remains at a high level, showing an average PCC value of 0.992 ± 0.001. However, when it comes to the completely overlapping input objects (r = 0), the PCC scores reduced to 0.795 ± 0.075. The PCC values quantified for the individual objects also showed that the axial planes closer to the front in the spatial sequence exhibit better imaging performance, revealing consistency with the previously shown output images.

To further explore the impact of varying $H_{\text{test}}$ on the QPI performance, we extended our analysis across an array of $H_{\text{test}}$ values {0.2, 0.4, 0.6, 0.8, 1, 1.2, 1.4, 1.6, 1.8}, all tested against the same diffractive QPI model trained with $H_{\text{tr}} = 0.6$, as shown in **Figure 3c**. It was found out that, the diffractive output QPI performance peaked at $H_{test} = 0.4$, with output PCC = 0.996 ± 0.001 for r = R, and PCC = 0.917 ± 0.021 for r = 0. Below this peak, when $H_{test} = 0.2$, a decrease in PCC was evident, demonstrating the challenge of resolving significantly lower phase contrast objects. In scenarios where the test thickness exceeded the training range ($H_{test} > H_{tr}$), our designs demonstrated some decrease in performance, especially as $H_{test}$ approached 1.8, where the PCC for r = R drops to 0.785 ± 0.139, and PCC for r = 0 drops to 0.646 ± 0.064. This decline can be attributed primarily to two factors. Firstly, significant phase contrast deviations between the training and testing expose the diffractive processor to unseen contrast levels, presenting generalization challenges. Secondly, the inherently linear nature of our diffractive processor, except for the intensity measurements at the output plane, faces approximation challenges under larger input phase contrast values due to the increased contributions of the nonlinear terms in the phase-to-intensity transformation task. Overall, our diffractive processor designs present decent generalization to various thicknesses and phase contrast values, very well covering $H_{test} \leq 1$ by using a fixed training thickness range parameter $H_{tr} = 0.6$ in the training stage.

**Impact of axial separation of input object planes on the multiplane QPI performance**

Beyond the lateral arrangement of the input phase objects, the axial distance separating these input planes is another crucial factor that influences the wavelength multiplexed QPI performance of our diffractive processors. To investigate this, we expanded our analysis of the output QPI performance by changing the input axial separation distance (Z) across {$128\lambda_m$, $64\lambda_m$, $32\lambda_m$, $16\lambda_m$}, as shown in **Figure 4**. Here, the testing thickness range parameter $H_{test}$ was fixed at 0.6. **Figure 4a** reveals that, when the axial distance decreases, the PCC values of the laterally overlapping phase inputs (r = 0) show a drop from 0.884 ± 0.016 at Z=$128\lambda_m$ to 0.802 ± 0.048 at Z=$16\lambda_m$. This decrease is expected due to the limited axial resolution of the diffractive QPI processor, leading to a degraded multiplane QPI performance for smaller Z distances. The output visualizations in **Figure 4b** corroborate these findings, displaying a noticeable decrease in image fidelity for multiplane QPI as Z decreases. Conversely, in scenarios with laterally separated inputs (r = 0.8R), the PCC values remained consistently high (around 0.993) even when the axial distance Z was reduced to $16\lambda_m$. When the input phase objects were partially overlapping (e.g., for r = 0.2R or 0.4R), the PCC values remained stable when Z decreased. This suggests that the diffractive processor maintains its effectiveness in phase reconstruction with laterally separated input phase objects regardless of the axial distance, Z. The output examples with varying distances further reinforce this conclusion, showing high-quality reconstructions across different axial separations. These observations underscore the diffractive processor's capability in multiplexed imaging of phase objects, especially when the inputs are not laterally overlapping.

**Crosstalk among imaging channels**

Ideally, our diffractive multiplane QPI processor should perform precise phase-to-intensity transformations for each input plane independently. However, accurately channeling the spatial information of individual object planes into their respective wavelength channels is challenging, as the features of the input objects positioned in the axial sequence can perturb the wave fields generated or modulated by the target object planes. This results in complex fields that, upon entering the diffractive processor, contain intermingled information from different object planes.

Consequently, information from one object plane can negatively impact the imaging process of another, leading to crosstalk among the imaging channels associated with different object planes. To delve deeper into the impact of this crosstalk among the channels, we conducted a numerical analysis by individually testing each input sample plane across all five wavelengths. By placing a phase object in one of the five object planes and leaving the remaining planes vacant, we could directly assess how the phase information from one input plane, corresponding to a specific output wavelength channel, affects other output channels. From the visualization of the output quantitative phase reconstructions shown in **Figure 5**, it was clear that when the inputs were laterally separated with r = R, the output images of the diffractive multiplane QPI processor at the target wavelength aligned well with the ground truth images and the signal leakage to the other wavelength channels was negligible. This result highlighted the diffractive processor's proficiency in handling and mitigating crosstalk between different wavelength channels. However, as the input separation distance decreased, for example, to r = 0.4, a noticeable crosstalk was observed across the different channels. This challenge became more pronounced when all the input objects were coaxially aligned at the center without any lateral separation (r = 0), resulting in more significant crosstalk as well as suboptimal quality of QPI reconstructions. These findings confirm the diffractive processor's capability to handle crosstalk effectively, while acknowledging its limitations when the input objects present a notable lateral overlap among different axial planes.

**Lateral resolution and phase sensitivity analysis**

To gain deeper insights into the diffractive multiplane QPI processor's capability to resolve phase images of input objects, we further investigated the lateral imaging resolution of our processor designs across different levels of input thickness range. To standardize our tests, we created binary phase grating patterns with a linewidth of $5.2\lambda_m$, and selected the testing thickness range parameter $H_{\text{test}}$ from {0.2, 0.6, 1} and the input lateral separation distance r from {0, 0.4R, R}, as shown in **Figure 6**. The results in **Figure 6a** and **Figure 6b** show that the diffractive QPI processors with $M = 5$ input planes effectively resolved the test phase gratings with a linewidth of $5.2\lambda_m$ for both r = 0 and r = 0.4R, even with thickness ranges that were different compared to the training thickness range, such as $H_{\text{test}} = 0.2 < H_{\text{tr}}$ or $H_{\text{test}} = 1 > H_{\text{tr}}$. In cases where r = 0, i.e., the test objects are positioned coaxially and exhibit complete lateral overlap, the processor can still resolve the grating patterns with $H_{\text{test}} = 1$ or $H_{\text{test}} = 0.6$, as shown in **Figure 6c**. However, at a thinner thickness or a smaller testing phase contrast level, e.g., $H_{\text{test}} = 0.2$, the resolution of diffractive QPI outputs became worse. The output examples revealed that the diffractive processor under r = 0 falls short in reconstructing the last two input planes (i.e., $P_4$ and $P_5$). Our analyses revealed that the diffractive multiplane QPI designs could clearly resolve spatial phase features with a linewidth of at least $5.2\lambda_m$ across all five input planes, particularly when the input phase object had a thickness range parameter $H_{\text{test}} > 0.2$.

**External generalization performance of wavelength-multiplexed diffractive processors for multi-plane QPI**

The diffractive multiplane QPI processors reported so far were trained on a dataset that included handwritten digits and grating-like spatial patterns. To further assess how well our diffractive multiplane QPI processors generalize to different types of spatial features, we conducted additional numerical analysis using Pap smear microscopy images. These images have significantly different spatial characteristics compared to our training dataset. In addition to this, we used various thickness range parameters ($H_{test}$) including {0.2, 0.4, 0.6, 0.8, 1, 1.2, 1.4, 1.6, 1.8} with an aim to examine the diffractive QPI processor's adaptability to new spatial features with previously unseen object thicknesses or phase contrasts, covering both $H_{test} > H_{tr}$ and $H_{test} < H_{tr}$. These blinded test results are showcased in **Figure 7b**, revealing a decent agreement between the diffractive multiplane QPI results and the corresponding ground truth images. We also calculated the image quality metrics across the entire Pap smear test dataset (see **Figure 7a**). The average PCC was calculated as 0.921 ± 0.004 when the testing thickness range matched the training condition, i.e., $H_{test} = H_{tr} = 0.6$. The QPI performance remained robust, with average PCC values of > 0.8 from $H_{test} = 0.2$ to $H_{test} = 1.4$, while starting to exhibit more degradation when $H_{test} > 1.4$. When $H_{test} = 1.8$, the average PCC dropped to 0.540 ± 0.113. Overall, these external generalization test results demonstrated that our diffractive multiplane QPI design is not limited to specific object types or phase features but can serve as a general-purpose multiplane quantitative phase imager for various kinds of objects.

**Output power efficiency of diffractive multiplane QPI processors**

All of our diffractive multiplane QPI processor designs presented so far were optimized without considering the output power efficiency, resulting in relatively low diffraction efficiencies, mostly lower than 0.1%. When the output power efficiency becomes a concern in a given diffractive processor design, an additional diffraction efficiency-related loss term[61,68,69] can be introduced to the training loss function to balance the tradeoff between task performance and signal-to-noise ratio. We used the same approach to achieve a balance between the QPI performance and the diffraction efficiency of the diffractive processor (see the Methods for details). In **Figure 8**, we present a comprehensive quantitative analysis of this tradeoff between the multiplane QPI performance and the output diffraction efficiency. For this comparison, we used two designs with r = R and 0.4R (as shown in **Figure S2, Supporting Information**), which initially exhibit output diffraction efficiencies of 0.113 ± 0.022% and 0.042 ± 0.004%, alongside PCC values of 0.993 ± 0.001 and 0.964 ± 0.021%, respectively - these are corresponding to diffractive designs trained *without* any diffraction efficiency penalty terms. Maintaining the same structural parameters and the same training/testing dataset, we retrained these wavelength-multiplexed diffractive multiplane QPI designs from scratch; this time, we incorporated varying degrees of diffraction efficiency penalty terms into the training loss functions, resulting in diffractive designs that demonstrated significantly enhanced output diffraction efficiencies. **Figure 8a** depicts the resulting PCC values of these new designs in relation to their output diffraction efficiencies. When compared to the original r = R design, these new designs showed an approximately **90-fold increase** in the output diffraction efficiency, which reached up to an efficiency of 10.2 ± 1.7%. This major output diffraction efficiency enhancement was achieved with a modest reduction in multiplane QPI performance, evidenced by PCC values decreasing to 0.842 ± 0.023. Similarly, compared to the original r = 0.4R design shown in **Figure S2, Supporting Information**, these new designs subjected to the diffraction efficiency penalty showcased an improved diffraction efficiency of up to 10.3 ± 0.7%, with a marginal decrease in the PCC values that reach up to 0.805 ± 0.042. Moreover, from the observation of the output examples in **Figure 8b**, the diffractive processors, even with enhanced output power efficiencies of >10%, still effectively reconstruct multiplane QPI images with a decent image quality. These results reveal that by properly incorporating an efficiency-related loss term into the optimization process, our wavelength multiplexed diffractive multiplane QPI processors can be optimized to maintain an effective balance between the QPI performance and the power efficiency, which is important for practical applications of the presented framework. This training approach to boost the output power efficiency was also used in our experimental proof of concept demonstration, which will be reported next.

**Experimental validation of a wavelength multiplexed diffractive multiplane QPI processor**

We conducted an experimental demonstration of our diffractive multiplane QPI processor using the terahertz part of the spectrum. As illustrated in **Figure 9a**, we created an input aperture to better control the illumination wavefront. The experimental configuration includes two input planes ($P_1$ and $P_2$), containing a phase-only object characterized by its thickness range parameter, empirically set as $H_{test} = \sim 0.7$. This setup serves as a proof-of-concept demonstration of our multiplane QPI system, wherein only one of two input planes contains a phase object at any given time. In our experiments, a diffractive multiplane QPI system composed of three phase-only dielectric diffractive layers ($L_1 - L_3$) was employed. This diffractive system converted the phase information of the input planes (axially separated by 20 mm) into an intensity distribution, captured at the output plane, where each illumination wavelength ($\lambda_1 = 0.8$ mm, $\lambda_2 = 0.75$ mm) was assigned to one axial plane performing QPI using phase-to-intensity transformations at each wavelength. Structural details of this experimental arrangement are provided in **Figure 9a** and the accompanying Methods section.

To optimize our experimental multiplane QPI design, we synthesized objects to train the diffractive design through deep learning. Our training dataset comprised 10,000 binary images of 4 x 4 pixels, with each image featuring two random pixels set to one and the remainder set to zero. These binary images were encoded into phase-only objects with a phase range of [0, $\alpha_{tr,w}\pi$], where the phase contrast parameter values $\alpha_{tr,w}$ reached 0.94 at $\lambda_1$ and 1 at $\lambda_2$. These phase contrast values were derived from the preset thickness range of [0, $H_{tr}\lambda_2$] where $H_{tr} = H_{test} = \sim 0.7$. Throughout the training process, for each iteration, one input plane was designated for the placement of the phase object, while the other was left vacant. The optimized phase profiles of the diffractive layers are displayed in the upper column of **Figure 9b**. After the training, the resulting diffractive layers were 3D printed, and images of the fabricated layers are showcased in the lower column of **Figure 9b**. After 3D assembly and alignment of these fabricated layers, we employed a terahertz source and a detector to record the intensity distribution at the output plane. Detailed schematics and photographs of this experimental setup are presented in **Figure S3, Supporting Information** and **Figure 9c**, respectively.

In the experimental phase, our system was subjected to eight distinct phase objects (never seen during the training) with the testing thickness range parameter set to $H_{test} = \sim 0.7$. These objects were equally divided between the two input planes that are axially separated by 20 mm, i.e., $\sim 25.8\lambda_m$, totaling four test phase objects per axial plane, and they were also fabricated using 3D printing. **Figure 9d** delineates the experimental output imaging results of the diffractive multiplane QPI processor, which align closely with our numerically simulated output patterns. The object phase profiles on both of the input planes were accurately transformed into intensity variations at the output plane, with each pixel clearly distinguishable and matching the expected ground truth phase profiles. These experimental results demonstrate the proof-of-concept capability of our diffractive design in conducting quantitative phase imaging across multiple planes using wavelength multiplexing.

## 3 DISCUSSION

In our results and analyses presented above, we have unveiled diffractive multiplane QPI processor designs utilizing wavelength multiplexing to encode the phase information of multiple input objects, which can be implemented through sequential imaging of different wavelength channels using, for example, a monochrome image sensor equipped with a spectral filter, each time adjusted to a unique wavelength; alternatively, a wavelength scanning light source can also be used for the same multiplane QPI. We would like to emphasize that our diffractive designs are not confined to multi-shot sequential image capture configurations, where the diffractive outputs for individual input planes are captured separately; our design framework can be further optimized to create a snapshot multiplane QPI system by devising the functionality of spectral filter arrays into the diffractive processor[66,70]. This functionality allows the multiplane phase signals at the diffractive processor's output to be partitioned, following a virtual filter-array pattern, enabling a monochrome image sensor to obtain signals from distinct object planes within a single frame. After a standard image demosaicing process, each QPI channel corresponding to a unique axial plane can be retrieved from a single intensity-only image.

It is crucial to highlight that our diffractive multiplane QPI design is tailored for a 3D stack of phase objects with weak scattering and absorption properties. This scenario meets the criterion for the first Born approximation[71], allowing the modeling of a 3D phase-only object using a discrete set of 2D phase modulation layers, which are assumed to be connected by free space propagation and approximately uniform illumination at each axial plane. The diffractive optical processor, due to its capacity for performing arbitrary complex-valued linear transformations between an input and output FOV[60], emerges as a viable approach for phase reconstruction and QPI under the first Born approximation. As one increases the lateral overlap among the axial planes that contain the phase-only input objects, the 3D QPI problem starts to deviate from the first Born approximation due to successive object-induced unknown wavefront distortions on the other axial planes where other unknown objects are located, which makes the problem nonlinear due to the interaction among the scattered fields that represent the object information at different planes. This physical cross-talk and the deviation from a linear coherent system approximation is at the heart of our QPI performance degradation observed for r = 0 when compared to the performance of r = R designs; the latter diffractive designs provide a better fit to the first Born approximation and the resulting fields at the output FOV of a diffractive QPI processor can be approximated as a linear superposition of the individual fields resulting separately from each axial plane. Having emphasized these points in relationship to the first Born approximation, we should also note that our numerical forward model does *not* make any such approximations and in fact precisely models object-to-object cross-talk fields for each case, taking all these nonlinear terms as part of its analysis and training/testing reported in this manuscript.

To further increase the performance of quantitative phase images and the spectral multiplexing factor (M), one would require a deeper diffractive architecture with more trainable degrees of freedom. Both theoretical analyses and empirical studies established earlier[57,59,60,62,63] have substantiated that increasing the total number of trainable diffractive features within a diffractive processor can improve its processing capacity and inference accuracy[63,72], also achieving significantly better diffraction efficiency at the output FOV. A particularly effective design strategy here involves increasing the number of layers rather than the number of diffractive features at each layer, which was proven to not only boost the diffraction efficiency but also achieve a more optimal utilization of diffractive features by enhancing optical connectivity between successive layers[60,69,73]. By increasing the number of diffractive layers (forming a deeper diffractive architecture), the performance of our wavelength multiplexed diffractive QPI processor can be further enhanced to perform the desired phase-to-intensity transformations more accurately across an even larger number of axial planes and also facilitate the multiplane QPI reconstructions with even a higher spatial resolution.

Notably, the presented multiwavelength diffractive processors maintain their accuracy in reconstructing quantitative phase images for multiple distinct planes irrespective of potential variations in the intensity of the broadband light sources used for illumination. Furthermore, these diffractive optical processors are not limited to the terahertz spectrum. By choosing suitable nano-fabrication techniques, including e.g., two-photon polymerization-based 3D printing[74–76], it is possible to scale these diffractive optical processors physically to operate across different segments of the electromagnetic spectrum, including visible and IR wavelengths. Such scalability and the passive nature of our diffractive processors pave the way for more efficient and compact on-chip phase imaging and sensing devices, promising a transformative impact for biomedical imaging/sensing and materials science.

## 4 EXPERIMENTAL SECTION

**Optical forward model of wavelength multiplexed diffractive QPI processors**

To numerically simulate a diffractive optical processor, each diffractive layer was treated as a thin optical element that modulates the complex field of the incoming coherent light. The complex transmission coefficient $t(x_q, y_q, z_l; \lambda)$ at any point $(x_q, y_q, z_l)$ on the $q^{\text{th}}$ diffractive feature of the $l^{\text{th}}$ layer is determined by the local material thickness, $h_q^l$, and can be described as:

$$t(x_q, y_q, z_l; \lambda) = \exp\left(\frac{-2\pi\kappa h_q^l}{\lambda}\right) \exp\left(\frac{-j2\pi(n - n_{\text{air}})h_q^l}{\lambda}\right) \quad (7).$$

In this equation, $n(\lambda)$ and $\kappa(\lambda)$ are the refractive index and extinction coefficient, respectively, of the chosen dielectric material at $\lambda$. These values correspond to the real and imaginary parts of the complex refractive index $\tilde{n}(\lambda) = n(\lambda) + j\kappa(\lambda)$. For the experimentally tested diffractive multiplane QPI processor, $n(\lambda)$ and $\kappa(\lambda)$ were set based on the measurements from a terahertz spectroscopy system[64]. As for the numerical analyzed diffractive multiplane QPI designs, $n(\lambda)$ was kept as the same, while $\kappa(\lambda)$ was set to 0. The thickness $h$ for each diffractive feature combines a constant $h_{\text{base}}$ and a variable/learnable part $h_{\text{learnable}}$, as shown in:

$$h = h_{\text{learnable}} + h_{\text{base}} \qquad (8),$$

where $h_{\text{learnable}}$ is the adjustable thickness part of each diffractive feature, constrained within the range [0, $h_{\text{max}}$]. In all the diffractive designs, including the numerical-simulated diffractive models and the experimentally validated diffractive model, $h_{\text{max}}$ is 1.4 mm, providing full phase modulation from 0 to $2\pi$ for the longest wavelength. The base thickness $h_{\text{base}}$, empirically set as 0.2 mm, provides the substrate (mechanical) support for all the diffractive features.

To simulate the light propagation of coherent optical fields in free space between the layers (including the input object planes, diffractive layers, and the output plane), we applied the angular spectrum approach[55]. The field at the $(l+1)^{\text{th}}$ diffractive layer, modulated by its transmittance function $t(x, y, z_{l+1}; \lambda)$, is given by:

$$u_q^{l+1}(x, y, z_{l+1}; \lambda) = t(x, y, z_{l+1}; \lambda)\mathcal{F}^{-1}\{\mathcal{F}\{u_q^l(x, y, z_{l+1}; \lambda)\}H_q^l(f_x, f_y, d_{\text{m}}; \lambda)\} \qquad (9).$$

Here, $\mathcal{F}\{\cdot\}$ and $\mathcal{F}^{-1}\{\cdot\}$ represent the 2D Fourier transform and its inverse operation, respectively. The transfer function $H_q^l(f_x, f_y, d_{\text{m}}; \lambda)$ of free-space propagation with a distance $d$ between two successive layers is given by:

$$H_q^l(f_x, f_y, d; \lambda) = \begin{cases} \exp\left\{\frac{j2\pi d}{\lambda}\sqrt{1 - (\lambda f_x)^2 - (\lambda f_y)^2}\right\}, & f_x^2 + f_y^2 < \frac{1}{\lambda^2} \\ 0, & f_x^2 + f_y^2 \geq \frac{1}{\lambda^2} \end{cases} \qquad (10),$$

where $f_x$ and $f_y$ represent the spatial frequencies along the x and y directions, respectively.

In our numerical simulations for all the diffractive designs demonstrated in this paper, we chose a spatial sampling rate for the simulated complex fields at a period of ~$0.44\lambda_{\text{m}}$. Similarly, the lateral size of the diffractive elements on each layer was selected at ~$0.44\lambda_{\text{m}}$. The axial distance between consecutive layers, including both the diffractive layers and the input/output planes, was set at $6\lambda_m$ for the numerical designs depicted in **Figure S2, Supporting Information**, and at $10\lambda_{\text{m}}$ for the design used in our experimental validations, as illustrated in **Figure 9a**.

**Numerical implementation of wavelength multiplexed diffractive multiplane QPI processors**

In our diffractive multiplane QPI processor design, multiple phase-only input objects are placed at different z-positions, where $z = z_1, z_2 \dots z_M$. Each object features a phase profile $\Psi_w(x, y; \lambda)$ with a consistent amplitude across the plane. The transmission through each of these phase objects/input planes is defined as:

$$t(x, y, z_w; \lambda) = e^{j\Psi_w(x,y;\lambda)} \qquad (11).$$

Initially, a broadband spatially-coherent source $s_w$ (or $u^0$) illuminates the front phase object plane. As the light propagates through axially stacked input planes, it is modulated by the phase objects and results in the complex field $u^l$ at each phase object plane. This field is calculated using the angular spectrum approach, post modulation by the object transmittance $t$, which can be expressed as $u^{l+1} = t \cdot \mathcal{F}^{-1}\{\mathcal{F}\{u^l\}H^l\}$. Finally, after traversing $M$ phase object planes, a cumulative multispectral complex field $i_w$ (or $u^M$) is formed, which contains the desired phase information of input objects.

The resulting field $u^M$ is then positioned at the entry point of the diffractive multiplane QPI processor. Consequently, the input field $u^M$ undergoes a sequence of diffractive layer modulations and secondary wave formations, as elaborated in the last subsection. This process ultimately results in a complex output field, denoted as $o_w(x, y) = u^{M+K}(x, y, z_K; \lambda_w)$, where K is the total number of the diffractive layers. Upon normalization with the reference signal ($Ref_w$) for each wavelength channel $w$, the resultant output QPI signals $\Phi_w$ can be obtained following Eq. (3).

For the diffractive QPI processor designs depicted in **Figure S2, Supporting Information**, both the input 2D FOVs distributed in the input 3D volume and the output FOV were designed to have a size of ~$73.5\lambda_m \times 73.5\lambda_m$. These FOVs are discretized into $14 \times 14$ pixels, with each pixel having dimensions of ~$5.25\lambda_m \times 5.25\lambda_m$. To ensure effective performance of the multiplane QPI task, every diffractive layer in this diffractive multiplane QPI processor contains $600 \times 600$ diffractive features, covering an area of ~$262.5\lambda_m \times 262.5\lambda_m$. These diffractive QPI processors used in our numerical analyses operate in the terahertz spectral range, i.e., $\lambda_1 = 0.9$ mm and $\lambda_M = 0.7$ mm. In the diffractive design used for our experimental validation, shown in **Figure 9a**, the input and output FOVs share identical dimensions of ~$24.8\lambda_m \times 24.8\lambda_m$. This space is divided into $4 \times 4$ pixels, resulting in each pixel being ~$6.2\lambda_m \times 6.2\lambda_m$. Each diffractive layer in this design includes $120 \times 120$ diffractive features, extending over an area of ~$49.5\lambda_m \times 49.5\lambda_m$. Here, the wavelengths for experimental validation were selected as $\lambda_1 = 0.8$ mm and $\lambda_2 = 0.75$ mm.

**Training data preparation and other implementation details**

For training our diffractive multiplane QPI processors, we assembled a dataset of 110,000 images, which can be divided into two categories: (1) 55,000 handwritten digit images from the original MNIST training set, and (2) 55,000 custom-designed images featuring a variety of patterns such as gratings, patches, and circles, each with unique spatial frequencies and orientations[67]. In the training phase, we generated each set of input objects by randomly choosing $M$ images from this dataset, with each image encoded into the phase channel of one of the $M$ object planes, thus creating an input object stack for the multiplane phase imaging task.

The numerical simulations and the training process for the diffractive multiplane QPI processors described in this study were carried out using Python (version 3.7.13) and PyTorch (version 2.5.0, Meta Platform Inc.). The Adam optimizer from PyTorch, with its default settings, was utilized. We set the learning rate at 0.001 and the batch size at 16. Our diffractive models underwent a 100-epoch training on a workstation equipped with an Nvidia GeForce RTX 3090 GPU, an Intel Core

i9-11900 CPU, and 128 GB RAM. The training time for a 10-layer diffractive multiplane QPI processor design, as seen in **Figure S2, Supporting Information**, was roughly 12 days, which is a one-time design effort.

**Details of the experimental diffractive multiplane QPI system**

Our diffractive multiplane QPI design was tested using a terahertz continuous wave (CW) system, as illustrated in **Figure S3, Supporting Information**. This setup involved a terahertz source comprising a Virginia Diode Inc. WR9.0M SGX/WR4.3x2 WR2.2 modular amplifier/multiplier chain (AMC), paired with a corresponding diagonal horn antenna (Virginia Diode Inc. WR2.2). At the AMC's input, a 10-dBm radiofrequency (RF) signal was introduced at 10.4166 or 11.1111 GHz (fRF1), which underwent a 36-fold multiplication, resulting in a 0.375 or 0.4 THz CW radiation output, equal to an illumination wavelength 0.8 or 0.75 mm, respectively. Additionally, for lock-in detection, the AMC's output experienced modulation with a 1-kHz square wave. Situated at ~10 mm from the horn antenna's exit plane, the input aperture was 1.6 mm wide. An XY positioning stage, comprising two Thorlabs NRT100 motorized stages, moved a single-pixel mixer (Virginia Diode Inc. WRI 2.2) to conduct a 2D scan of the output intensity distribution, with a step size of 0.8 mm. The detector also received a 10-dBm RF signal at 11.1111 or 10.4166 GHz (fRF2) as the local oscillator, down-converting the output frequency to 1 GHz. This down-converted signal then passed through a low-noise amplifier (gain: 80 dBm) and a KL Electronics 3C40-1000/T10-O/O bandpass filter at 1 GHz (+/-10 MHz), reducing noise from undesirable frequency bands. After a linear calibration with an HP 8495B tunable attenuator, the signal was relayed to a Mini-Circuits ZX47-60 low-noise power detector. The lock-in amplifier (Stanford Research SR830) then processed the detector's output voltage, using the 1-kHz square wave as a reference for linear scale calibration.

For the fabrication of the diffractive multiplane QPI system depicted in **Figure 9b**, an Objet30 Pro 3D printer by Stratasys was employed to print the diffractive design and the input aperture. To ensure alignment with our optical forward model for the experimental diffractive design, a 3D-printed holder was fabricated using the same printer. This holder facilitated the precise positioning of both the input aperture and the printed diffractive layers, securing their accurate 3D assembly.

**Supporting Information:** This file contains:

- Training loss function and image quality metrics
- Figures S1-S3.

**Figures**

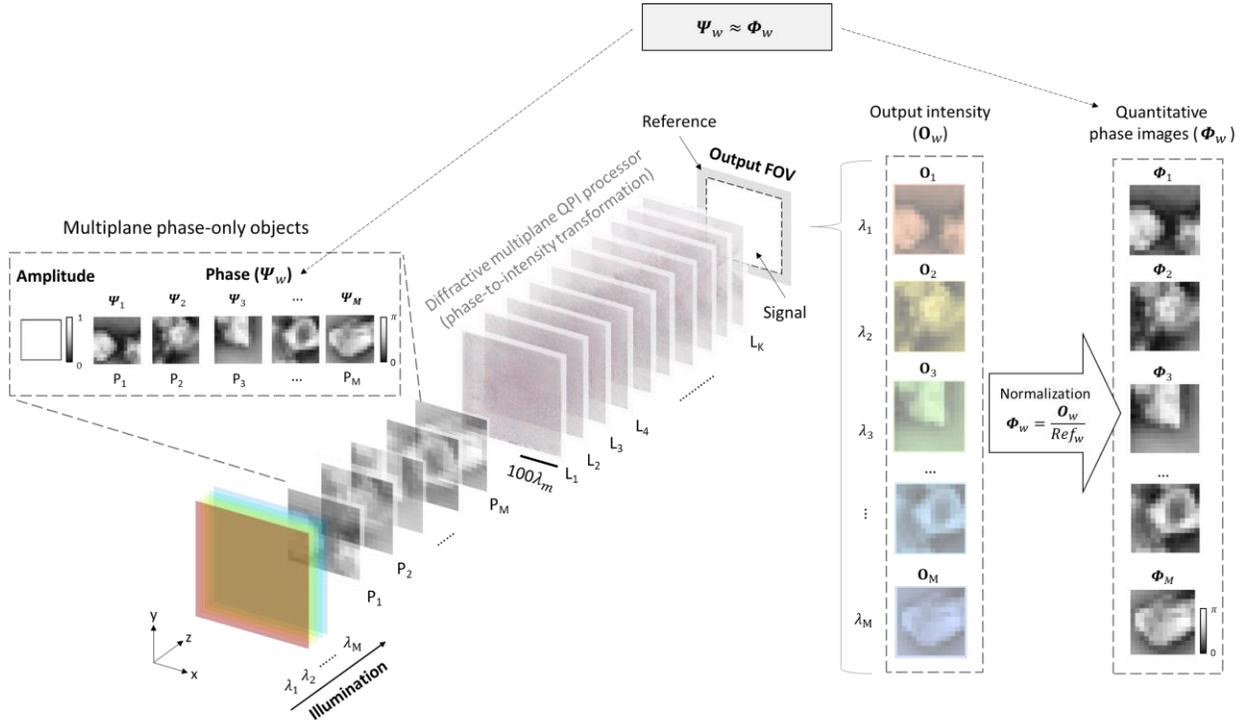

**Figure 1. Schematic and working principle of multiplane quantitative phase imaging (QPI) using a wavelength-multiplexed diffractive processor.** Illustration of a wavelength-multiplexed diffractive multiplane QPI processor. The diffractive QPI processor is composed of *K* diffractive layers, which are jointly optimized using deep learning to simultaneously perform phase-to-intensity transformations for *M* phase-only objects that are successively positioned along the axial direction (z), while also routing QPI signals of these objects to the designated wavelength channels at the same output FOV.

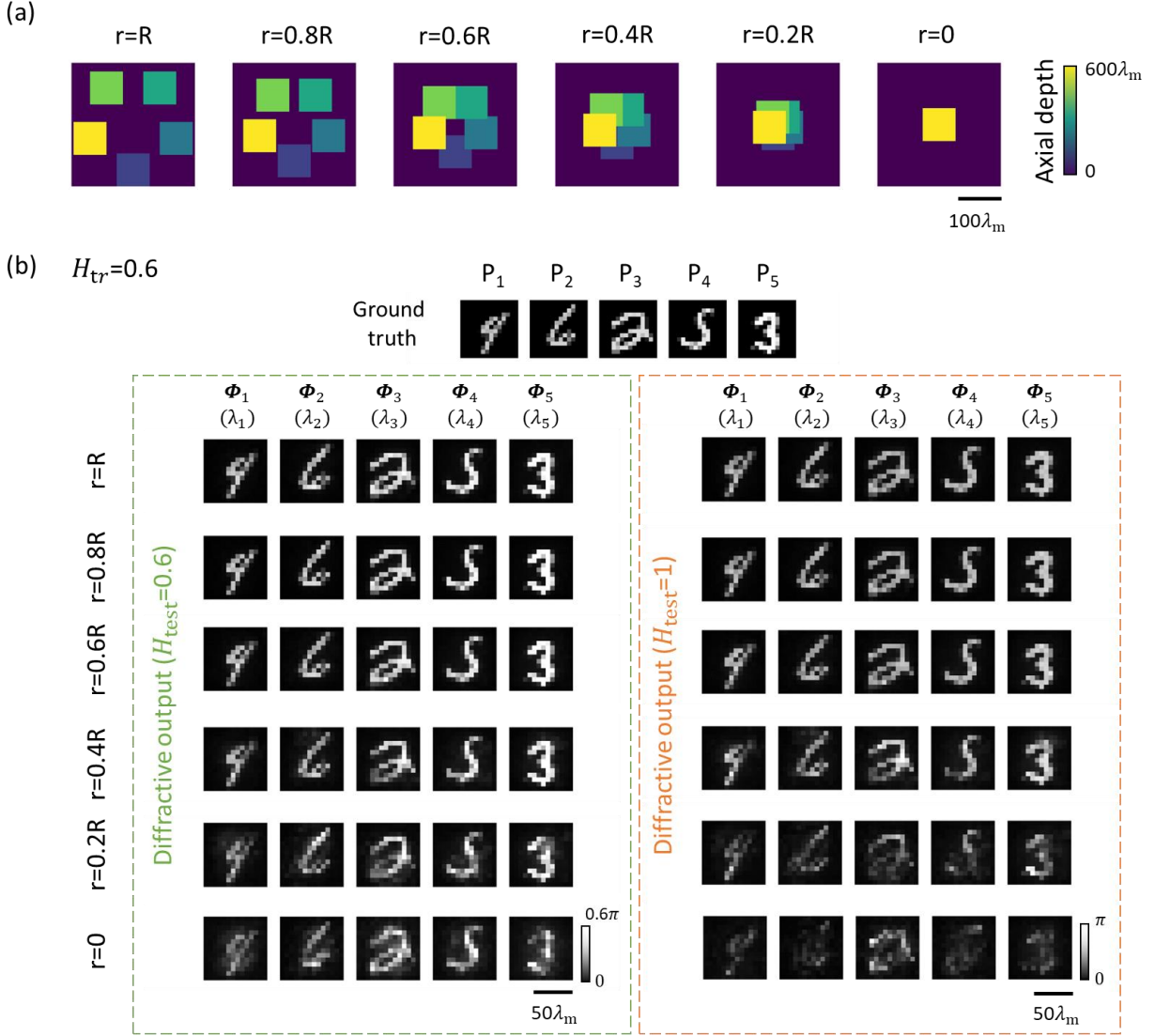

**Figure 2. The lateral separation settings of the input objects and the blind testing results of the diffractive multiplane QPI processors. a,** Input volume visualization for six different diffractive designs under different input lateral separation distances (r) spanning {0, 0.2R, 0.4R, 0.6R, 0.8R, R}. The colormap represents the input object distribution in the axial range. **b,** Examples of the blind testing results for multiplane QPI using six different diffractive processors under different input lateral separation distances (r).

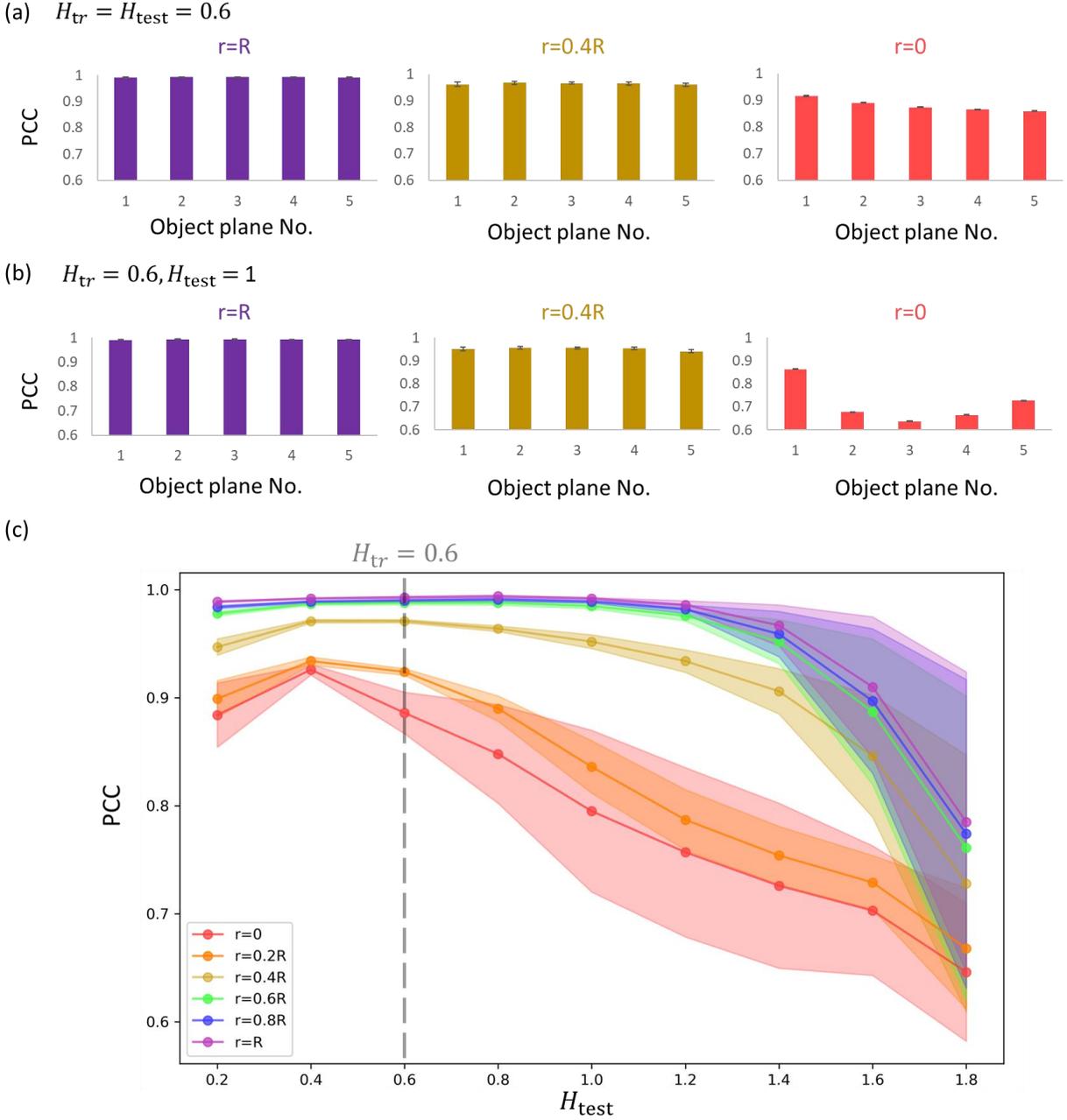

**Figure 3. Impact of the input lateral separation and the input object thickness on multiplane QPI performance. a,** PCC values of the resulting multiplane QPI measurements with $H_{\text{test}} = 0.6$ under different input lateral separation distances (r) spanning {0, 0.2R, 0.4R, 0.6R, 0.8R, R}. **b,** same as a, except for $H_{\text{test}} = 1$. **c,** Average PCC values of the resulting multiplane QPI measurements as a function of $H_{\text{test}}$. These six curves refer to the blind testing performances of the six diffractive processors trained under different input lateral separation distances (r).

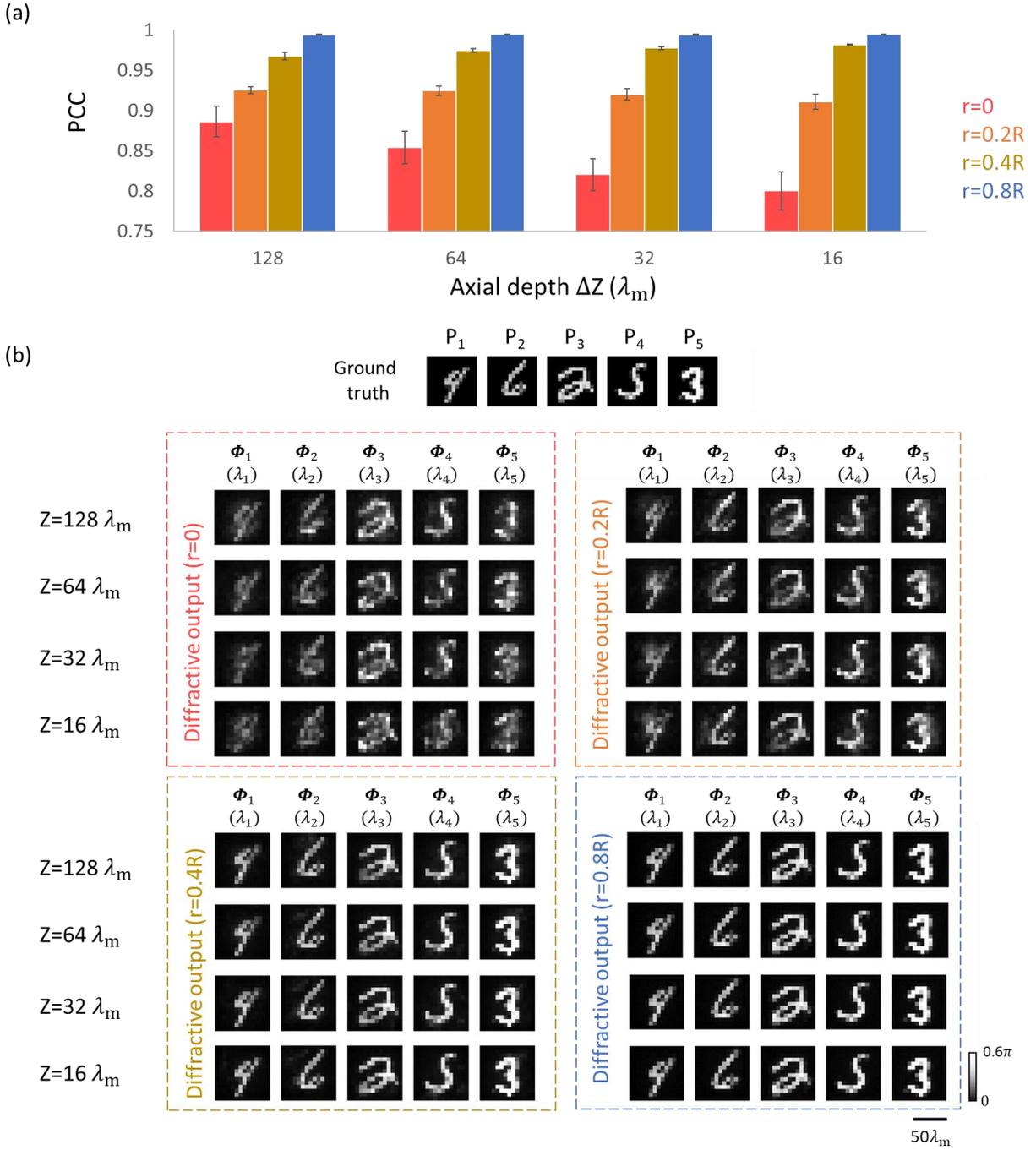

**Figure 4. Impact of the input axial separation on the output multiplane QPI performance. a,** Average PCC values of the diffractive multiplane QPI processor outputs with different input lateral separation distances (r) covering {0, 0.2R, 0.4R, 0.8R} and different input axial separation distances (Z) covering {$128\lambda_m$, $64\lambda_m$, $32\lambda_m$, $16\lambda_m$}. **b,** The corresponding output examples of the diffractive multiplane QPI results.

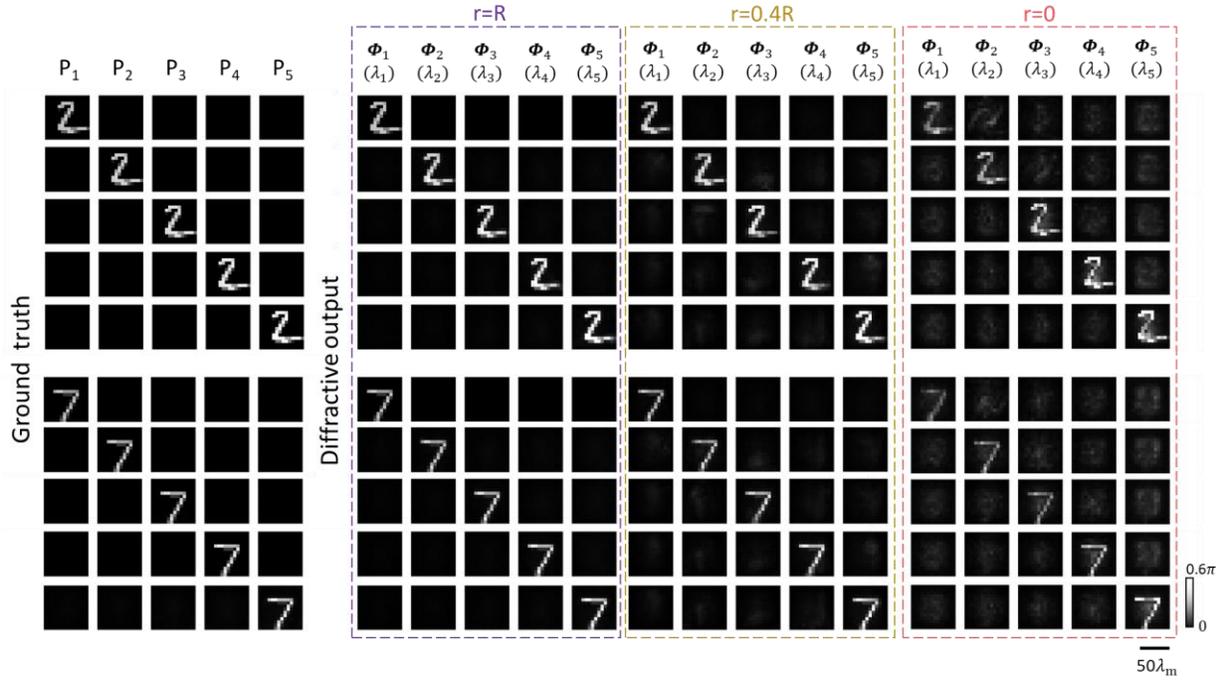

**Figure 5. Crosstalk analysis of multiplane QPI under different input lateral separation distances.** Output image matrix demonstrating the crosstalk from one input plane to the output wavelength channels, represented by the off-diagonal images. Each row corresponds to a set of input (ground truth) phase objects alongside the resulting diffractive output images. The diagonal images represent the diffractive output images at the target wavelengths.

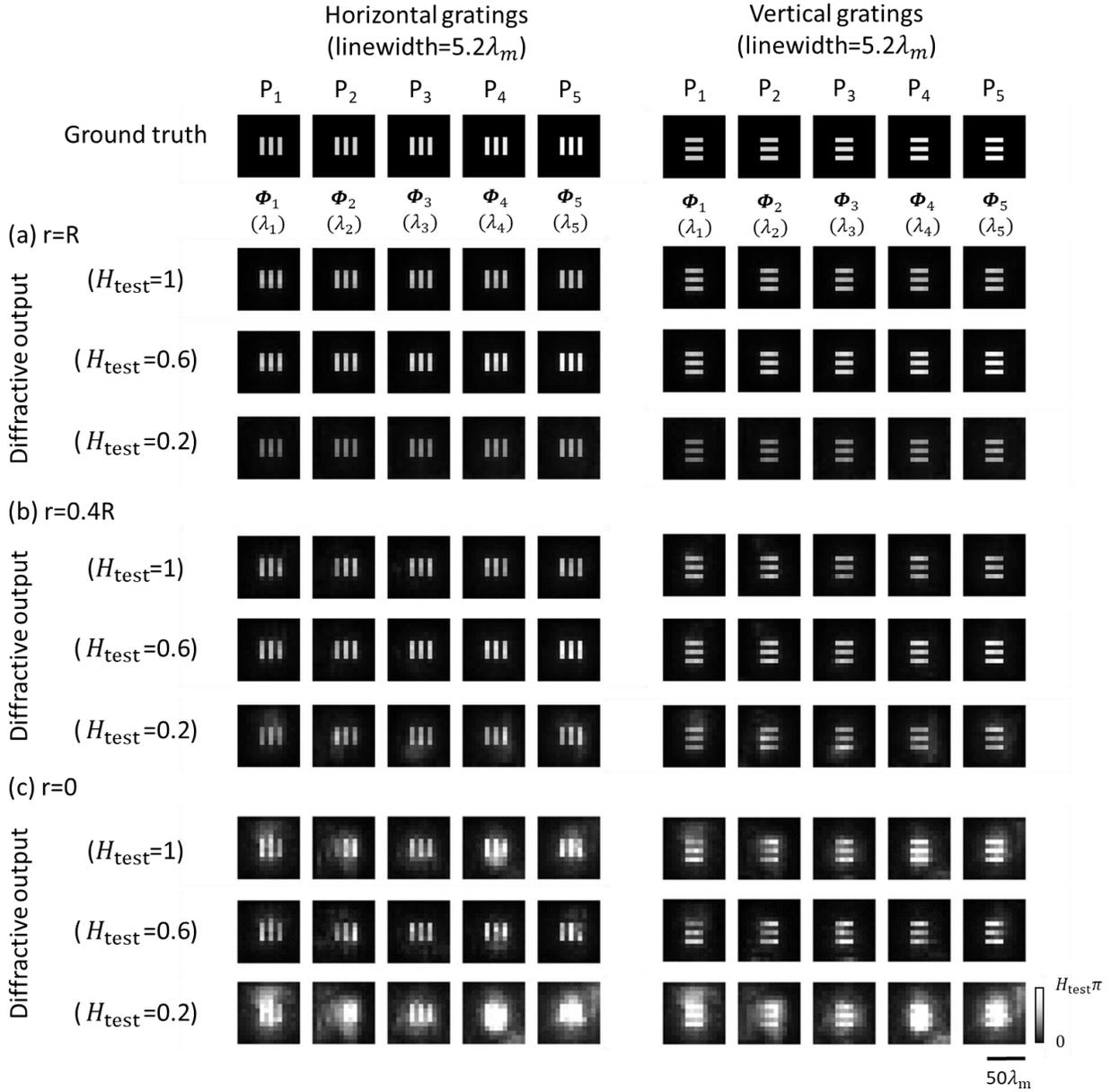

**Figure 6. Lateral resolution and phase sensitivity analysis for the diffractive multiplane QPI processor designs. a.** Images of the binary phase grating patterns encoded within the phase channels of the input object, along with the r = R diffractive processor's resulting output QPI signals ($\Phi_w$) at the target input plane. The grating has a linewidth of $5.2\lambda_m$, and the thickness range parameter ($H_{test}$) of the input phase object is selected from {0.2, 0.6, 1}. **b, c.** same as **a**, except for r = 0.4R in **b**, and for r = 0 in **c**.

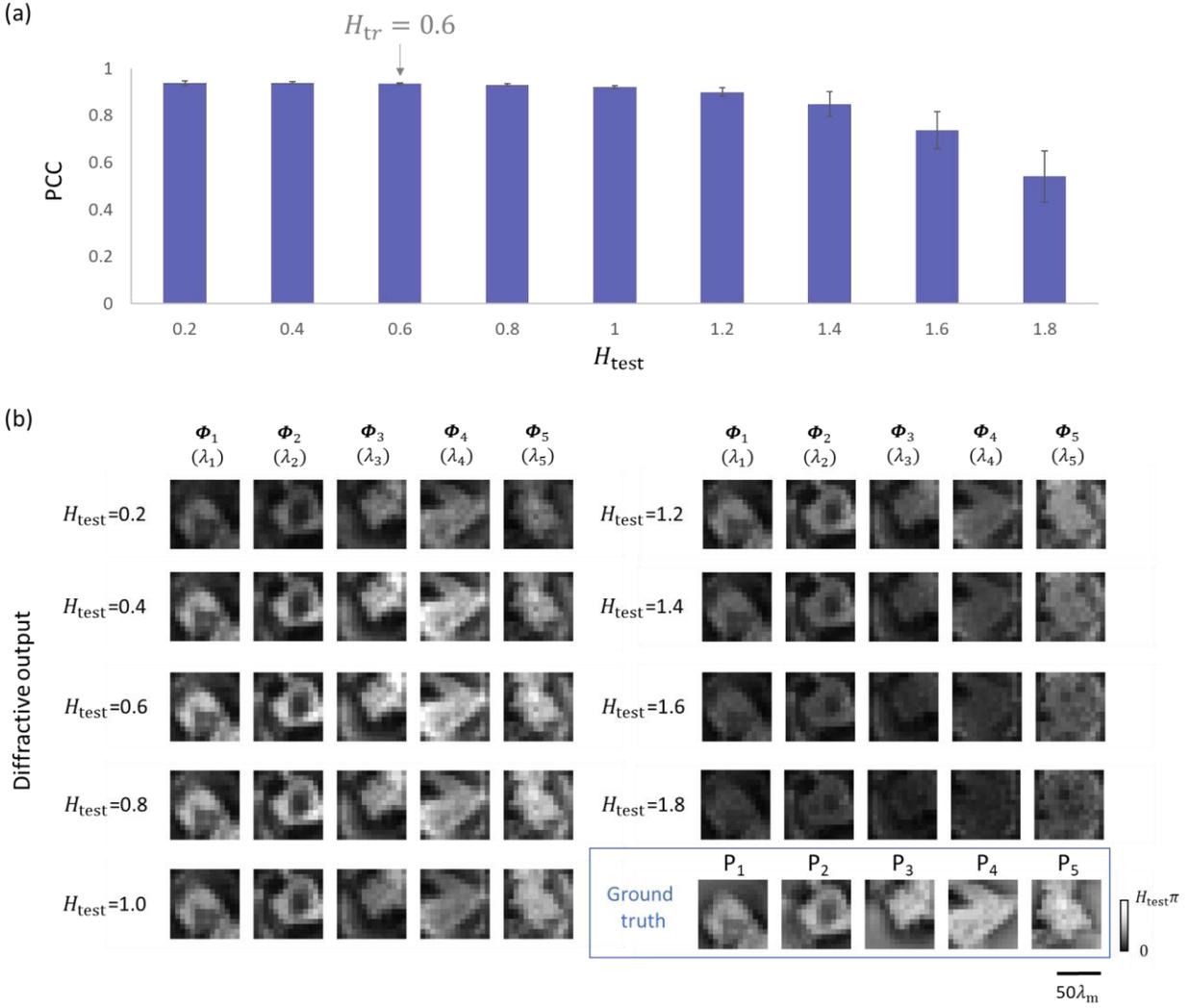

**Figure 7. Results for testing the external generalization performance of the r = R diffractive multiplane QPI processor design using blind testing images from a new dataset composed of Pap Smear images. a,** PCC values of the diffractive multiplane QPI processor outputs as a function of the input thickness range. **b,** Examples of the ground truth phase images at different input planes, which are compared to their corresponding diffractive QPI output images.

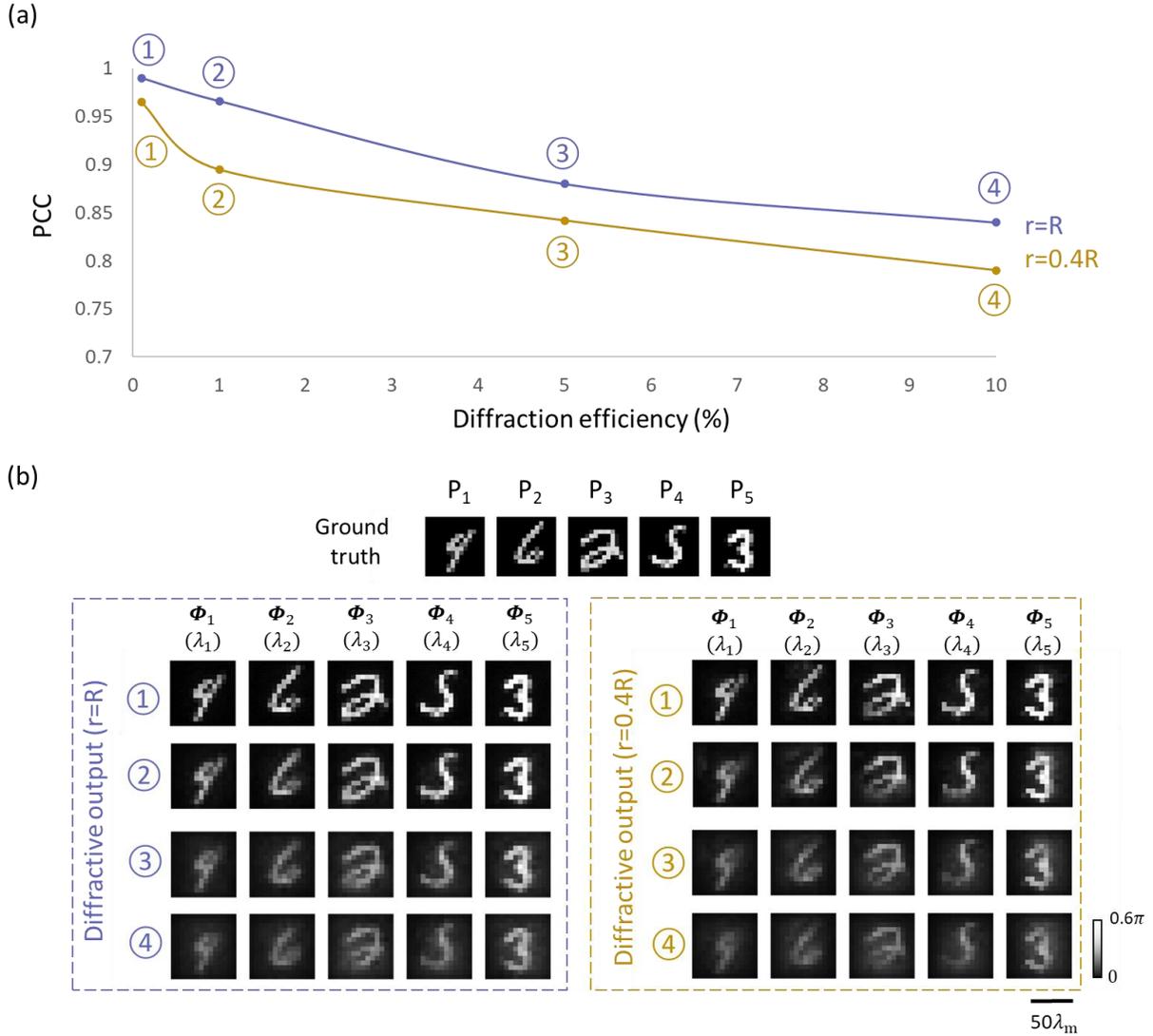

**Figure 8. Analysis of the tradeoff between the imaging performance and the output diffraction efficiency of diffractive multiplane QPI processors. a,** The PCC values of the diffractive multiplane QPI outputs with various levels of diffraction efficiency penalty, plotted as a function of the output diffraction efficiency values. Two sets of diffractive QPI designs using r = R and r = 0.4R were trained and blindly tested. Specifically, purple markers (①, ②, ③ and ④) depict different r = R designs where $\beta_{Eff} = 0$ was used for ① and $\beta_{Eff} = 100$, $\eta_{thresh} = 1\%$, 5%, 10% were used for ②, ③ and ④, respectively, in the training loss function (see Eqs. (13-14)). Gold markers (①, ②, ③ and ④) represent their counterparts using r = 0.4R. **b,** Visualization of the diffractive output fields produced by diffractive QPI processor designs with different input lateral separation distances and various levels of diffraction efficiency-related penalty term.

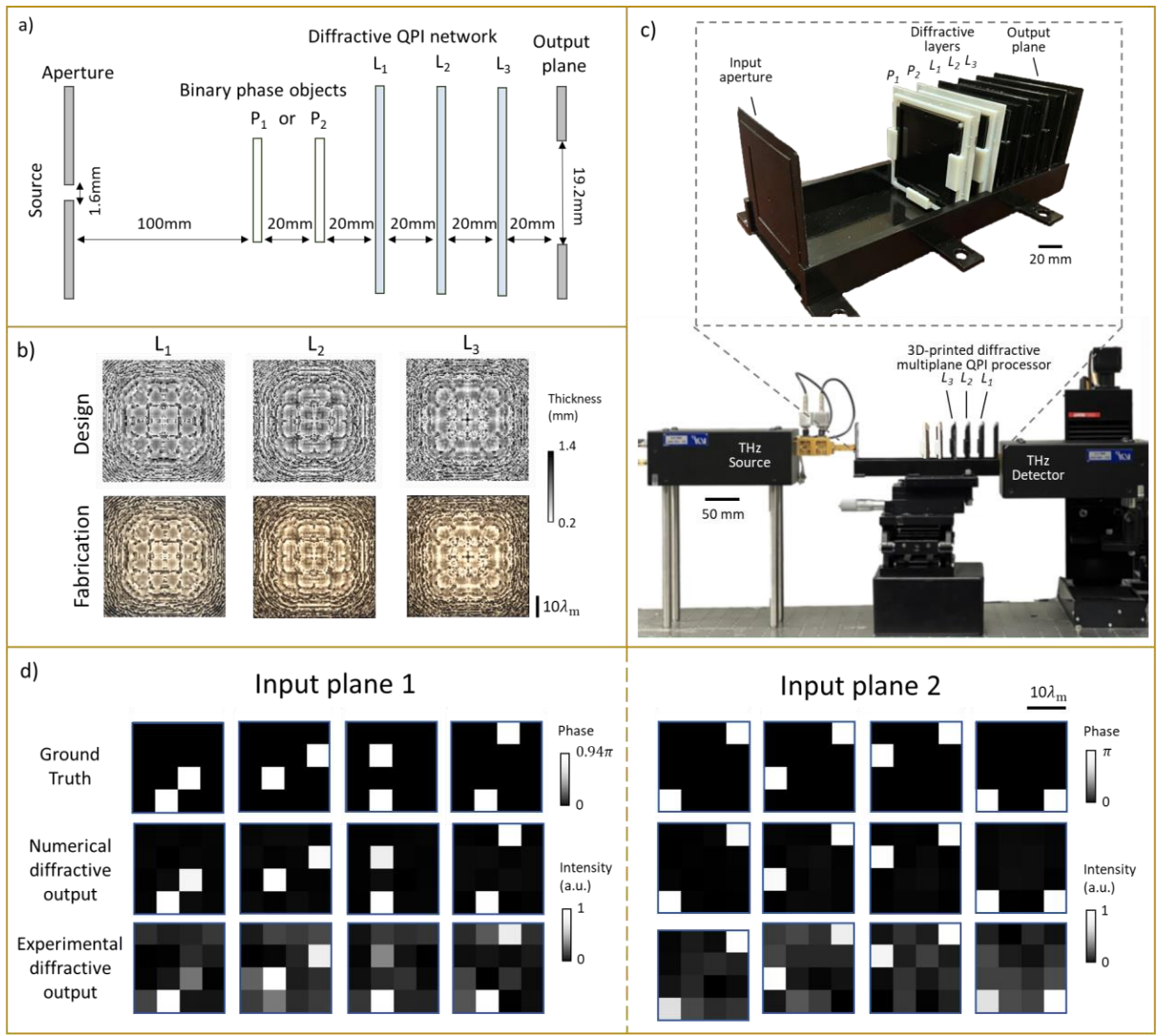

**Figure 9. Experimental set-up and validation of the diffractive multiplane QPI processor for phase-to-intensity transformations. a,** Illustration of a diffractive multiplane QPI processor composed of three diffractive layers ($L_1$, $L_2$, $L_3$) to perform QPI operation on multiplane phase objects. **b,** Thickness profiles of the optimized diffractive layers (upper row) and the photographs of their fabricated versions using 3D printing (lower row). **c,** Photographs of the experimental set-up, including the fabricated diffractive QPI processor. **d,** Numerically simulated and experimentally measured intensity patterns at the output plane, compared with the ground truth input objects, successfully demonstrating experimental phase-to-intensity transformations.